\documentclass[aps,prl,twocolumn,amsmath,amssymb,superscriptaddress,citeautoscript,longbibliography,floatfix]{revtex4-1}
\newcommand{\angstrom}{\textup{\AA}}
\usepackage[allow-number-unit-breaks]{siunitx}
\usepackage{graphicx}
\usepackage{dcolumn}
\usepackage{bm}
\usepackage{color}
\usepackage{bbding}
\usepackage{pifont}
\usepackage{url} 
\newcommand{\xmark}{\ding{55}}
\usepackage{mathrsfs}
\usepackage{amsfonts}
\usepackage{multirow}
\usepackage{array}
\usepackage{natbib}
\usepackage{epsf}
\usepackage{epsfig}
\usepackage{epstopdf}
\usepackage{amsmath}
\usepackage{hyperref}
\usepackage{changes}

\hypersetup{
    colorlinks=true,
    linkcolor=blue,
    filecolor=cyan,      
    citecolor=blue,
    urlcolor=blue,
    }

\begin{document}

\title{Nonrelativistic Optically-Induced Spin Torque in Noncollinear Magnets}

\author{Zhiyuan Li}
\affiliation{Center for Alloy Innovation and Design, State Key Laboratory for Mechanical Behavior of Materials, Xi'an Jiaotong University, Xi'an, 710049, China}

\author{Jian Zhou}\email{jianzhou@xjtu.edu.cn}
\affiliation{Center for Alloy Innovation and Design, State Key Laboratory for Mechanical Behavior of Materials, Xi'an Jiaotong University, Xi'an, 710049, China}

\begin{abstract}
Manipulating spin using contactless and non-invasive optical approaches is one of the forefront topic in the modern spintronic science and technology. Prior studies on optically-induced spin reorientation usually scales with spin-orbit coupling strength, setting a limitation on composing heavy elements. In this work, we extend such a limit into nonrelativistic regime by exploring noncollinear magnetic configuration. Using a general spin group-theoretical investigation, we show how the symmetric operators in both spin and lattice spaces control the spin torque and quantum geometry of Bloch wavefunction--including both quantum metric and Berry curvature. We describe their site-resolved spin symmetries and unravel their selection rules. Furthermore, we adopt a Kagom\'e lattice model and a realistic material family to showcase the local spin torque under light irradiation. Since the nonrelativistic spin splitting usually spans on the order of electron-volt, significantly larger than typical spin-orbit induced splitting, the nonrelativistic light-induced spin torque is concomitantly much stronger. This work extends the previous opto-spintronics from relativistic to nonrelativistic limit, significantly broadening its realization platform.
\end{abstract}

\maketitle
\textit{Introduction.}---The manipulation of spin direction in magnetic materials is one crucial and essential process in the spintronic information technology \cite{Zutic04RevModPhys.76.323,Baltz18RMP,Han23NM,Nemec18NP,Yang21NatRevPhyss42254-021-00302-9,Avsar20RevModPhys.92.021003,Bai24AFM}. Of various strategies such as thermal approaches \cite{Geranton15PhysRevB.91.014417}, magnetic field \cite{Hur04Nature}, and electric field/current \cite{Haney10PhysRevLett.105.126602,Zelezny14PhysRevLett.113.157201,Han24SA,Han24NatPhys,Reimers23NC,Chen25PRLzm5y-vy41,Xiang25PhysRevLett.134.116301,Cao25PRB54qq-2qgh,Qiang26PRL4kmy-59l9}, light-induced (de-)magnetization and reorientation (spin torque) has attracted special interest \cite{Kirilyuk10RMP,Pervishko15PRB,Juraschek20PRR,Kibis22PRA,Watzel22PRL,Kahana24SAado0722,Wu26SAadv9616,Xu21PRB,Xue24ACSNano,Wu24NCs41467-024-47128-4,Zhou25PhysRevLett.134.176902} due to its noncontacting and highly-tunable remote control characters. In the perturbative quantum mechanical framework, the optically-induced spin torque (OIST) is frequency-dependent second order version of spin-orbit torque (SOT) process \cite{Freimuth14PhysRevB.90.174423,Wimmer16PhysRevB.94.054415,Belashchenko19PRM.3.011401,Geranton15PhysRevB.91.014417,Smejkal17PRL,Zelezny17PhysRevB.95.014403,Bodnar18NC}, \textcolor{black}{which describes how an electric field/current induces a spin torque in ferromagnetic and/or antiferromagnetic materials.} While the SOT could occur in a single phase material, it is preferable to be applied in a heterojunction between heavy-metal (with sufficient spin-orbit coupling, SOC) and the magnetic system. From the symmetric point of view, the SOT requires inversion symmetry ($\bar{E}$) broken, while the OIST or spin accumulation--as second order responses--can be applied in centrosymmetric systems \cite{Xu21PRB}. This facilitates a broader material platforms for such a spin control.

Prior works on opto-magnetic interactions rely on the relativistic framework with finite SOC, akin to the conventional SOT. This greatly limits its material family which should contain heavy elements. Recent years have seen an increasing attention on the spin texture in the nonrelativistic (NR) limit, such as spin splitting in electronic band dispersion \cite{Krempasky24Nature,Fedchenko24SA,Lee24PRL}, charge and spin transport \cite{Fang24PRL,Zhu25NCs41467-025-60128-2,Yoshida26PRL463y-q7lt,cuono26arxiv.chargespinphotogalvaniceffects}, magneto-optical reflection and refraction \cite{Pan26PRLq8ym-l2zt,Iguchi25PRRnnz3-tq7y}, anomalous and spin Hall conductance \cite{Feng22NE,Reichlova24NC}, SOT and current-induced Edelstein effect \cite{Gonzalez24NCs41467-024-51565-6}, especially with the discovery of unconventional antiferromagnetic (dubbed altermagnetic) systems \cite{Ma21NC,Smejkal22PRX,Smejkal22PRX-2,Jiang25NatPhyss41567-025-02822-y} and the associated spin group theory \cite{Brinkman66PRSA,Liu22PRX,Xiao24PhysRevX.14.031037,Cheong24npjQM}. From these studies, the large spin splitting nature (lies on the order of $0.1-1\,\rm eV$) in NR-regime usually guarantees stronger spintronic responses, if symmetry-allowed, than the SOC-induced feature. \textcolor{black}{These advances imply that a NR-OIST could potentially overcome the restriction to heavy elements, and one may realize a contactless optical control on the spin order.} \textcolor{black}{NR spin torques have been studied mainly in the current-driven regime, such as an intrinsic electric field-induced nonadiabatic spin-transfer torque in smooth magnetic textures \cite{PhysRevB.92.224426} and noncollinear magnets \cite{Gonzalez24NCs41467-024-51565-6}. However, their second-order optical responses in noncollinear magnets is still missing, and its general NR symmetry theories remain to be investigated.} 

In this work, we aim to explore such a challenge by adopting a group-theoretical analysis and predict the NR-OIST using a minimum tight-binding model and first-principles calculations. We use \textcolor{black}{density matrix approach} and derive the perturbative response expression of OIST. We then adopt spin group analysis and show that the NR-OIST only arises in noncollinear magnetic systems. We enumerate how they govern the NR-OIST by investigating all site-resolved spin point groups (SPGs) and their respective constraints on spin torque and Bloch band quantum geometry. Furthermore, we adopt our theory in Kagom\'e lattice and realistic $\rm Mn_3Ge$ ($\rm Mn_3Sn$) materials. Our calculations suggest that a modest light field could induce observable spin torque on the magnetic sites, with the effective magnetic field on the order of $10\,\rm T$. The NR-OIST response is in general much stronger than the SOC-induced relativistic OIST by $1-2$ orders of magnitude.

\textit{Nonrelativistic optically-induced spin torque.}---We start from a general NR Hamiltonian
\begin{equation}\label{eq:ham}
    H=H_0+\sum_iH_{\mathrm{ex}}^i=H_0+\sum_iJ\hat{\bm s}_i\cdot\bm\sigma,
\end{equation}
where $H_0$ is spin-independent and $J$ denotes a uniform exchange field. $\hat{\bm s}_i$ is a unit vector along local spin at site-$i$. Under a perturbative field, the spin torque is defined as a commutator with the Hamiltonian $\bm T^{i}=\frac{d\bm\sigma}{dt}=\frac{1}{i\hbar}[\bm\sigma,H_{\mathrm{ex}}^i]=J\hat{\bm s}_i\times\bm\sigma$ which is always perpendicular to the equilibrium spin \cite{Wimmer16PhysRevB.94.054415,Belashchenko19PRM.3.011401,Geranton15PhysRevB.91.014417,Freimuth15PhysRevB.92.064415}. This measures a dynamical rate rather than the spin accumulation, \textcolor{black}{hence is a $\mathcal{T}$-even operator. Here, the $\mathcal{T}$-even nature indicates that the torque value remains unchanged if all spins are flipped.} Under an optical electric field $\vec{\mathcal{E}}(\omega)=\mathrm{Re}(\bm Ee^{i\omega t})$ with angular frequency of $\omega$, we consider its second order real symmetric [$\mathrm{Re}(E_{\alpha}E_{\beta}^*)$] and imaginary antisymmetric parts [$\mathrm{Im}(E_{\alpha}E_{\beta}^*)$]. \textcolor{black} {They correspond to linearly and circularly polarized light (LPL and CPL) effect, namely, $T_{\gamma}^{\mathrm{L},i}=\sum_{\alpha\beta}\chi_{\alpha\beta}^{\gamma,i}(\omega)\mathrm{Re}(E_{\alpha}E_{\beta}^*)$ and $T_{\gamma}^{\mathrm{C},i}=\sum_{\alpha\beta}\xi_{\alpha\beta}^{\gamma,i}(\omega)\mathrm{Im}(E_{\alpha}E_{\beta}^*)$.} Both these susceptibility functions arise from the quantum geometry tensor $Q^{\alpha\beta}=g^{\alpha\beta}+\frac{i}{2}\Omega^{\alpha\beta}$,
\begin{equation}\label{eq:oist}
    \begin{split}
        \chi_{\alpha\beta}^{{\gamma},i}(\omega)=&\frac{\tau e^2\pi V_{\mathrm{c}}}{\hbar^2}\int[d\bm k]\sum_{n,m}f_{nm}\Delta T_{nm}^{\gamma,i}g_{mn}^{\alpha\beta}\delta(\omega-\omega_{mn}),\\
        \xi_{\alpha\beta}^{{\gamma},i}(\omega)=&\frac{\tau e^2\pi V_{\mathrm{c}}}{2\hbar^2}\int[d\bm k]\sum_{n,m}f_{nm}\Delta T_{nm}^{\gamma,i}\Omega_{mn}^{\alpha\beta}\delta(\omega-\omega_{mn}).
    \end{split}
\end{equation}
The detailed derivation under length-gauge can be seen at the End Matter and Supplmental Material \textcolor{black}{(SM)} \cite{supp,Aversa95PhysRevB.52.14636,Sipe20PRB,Liu25PRLnsky-zmzp,Bodnar18NC,Freimuth14PhysRevB.90.174423,Tao24PRL,Gonzalez24NCs41467-024-51565-6,Liu22PRX,Kresse93PRB,Kresse96PRB,Perdew96PRL,Perdew08PRL,Blochl94PRB,Monkhorst76PRB,Mostofi14CPC}. Here, $f_{nm}=f_n-f_m$, $\omega_{mn}=\omega_m-\omega_n$, and $\Delta\bm T_{nm}^i=\langle n|\bm T^i|n\rangle-\langle m|\bm T^i|m\rangle$ measure the Fermi-Dirac occupation, eigenfrequency, and spin torque (at site-$i$) difference between band $n$ and $m$. The quantum geometry tensor components \cite{Yu25npjQM,Liu25NSRnwae334,Verma26NatPhys,Ghosh24SAado1761} are defined by position matrix $r_{mn}^{\alpha}=\langle m|r^{\alpha}|n\rangle$, giving $g_{mn}^{\alpha\beta}=(r_{mn}^{\alpha}r_{nm}^{\beta}+r_{mn}^{\beta}r_{nm}^{\alpha})/2$ and $\Omega_{mn}^{\alpha\beta}=i(r_{mn}^{\alpha}r_{nm}^{\beta}-r_{mn}^{\beta}r_{nm}^{\alpha})$. $V_{\mathrm{c}}$ is the volume of unit cell and $\tau$ denotes the phenomenological carrier lifetime. All quantities are $\bm k$-dependent and the integral is performed in the whole $D$-dimensional BZ $[d\bm k]=\frac{d^D\bm k}{(2\pi)^D}$. \textcolor{black}{Here, $\chi$ and $\xi$ are akin to the linear and circular injection current susceptibility} \cite{Sturman20PU,Zhao26adma.202517783}. The non-diagonal elements representing spin-shift contributions (similar as the shift current) are omitted as they are order of magnitude weaker \cite{Fregoso22PRB,Zhou25PhysRevLett.134.176902,Sarkar_2025NJPae1868} (see SM \cite{supp} for details).

\textit{Symmetry analysis.}---Now we perform a group-theoretical analysis to investigate the symmetry constraints of the above OIST responses. Since we are focusing on the NR regime, we adopt spin group to conduct the analysis. In the spin space group framework, a general symmetric operation takes the form of $[R_s||R_l|\bm t]$ where $R_s$ and $R_l|\bm t$ ($\bm t$ denoting fractional lattice translation) act on spin and lattice spaces separately. In our analysis, the fractional lattice translation does not affect the quantum geometry and spin torque, which can be omitted here. Hence, we mainly use SPG (denoted as $\mathcal{G}$) instead of spin space group. The symmetry transformation of OIST can be individually analyzed for spin torque vector $\bm T$ and quantum geometric tensors $g^{ij}$ and $\Omega^{ij}$. The spin torque $\bm T$ can be viewed as an axial vector in spin space, hence it is invariant when pointing along a rotation axis or vertical to a reflection mirror, namely, $C_{n\alpha}T_{\alpha}=\bar{C}_{n\alpha}T_{\alpha}=T_{\alpha}$, $M_{\alpha}T_{\alpha}=\bar{M}_{\alpha}T_{\alpha}=T_{\alpha}$, while it flips sign under $C_{2\alpha}T_{\beta}=\bar{C}_{2\alpha}T_{\beta}=-T_{\beta}$ and $M_{\alpha}T_{\beta}=\bar{M}_{\alpha}T_{\beta}=-T_{\beta}$ ($\alpha$ and $\beta$ are orthogonal directions and the bar indicates spin-space inversion).

\begin{figure}[bt]
    \centering
    \includegraphics[width=0.48\textwidth]{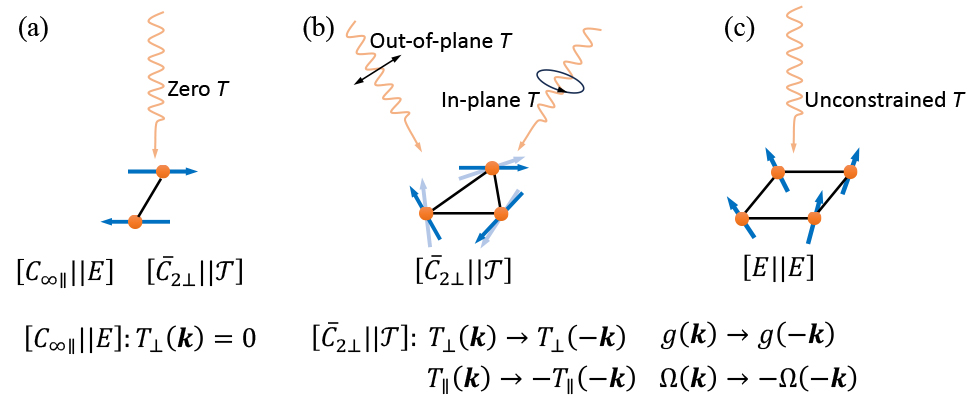}
    \caption{Spin-only group effect on the nonrelativistic spin torque. (a) Collinear spin (blue arrows) pattern forbids the NR spin torque. (b) Coplanar spin structure that supports linearly (circularly) polarized light induced out-of-plane (in-plane) spin torque. (c) A general noncoplanar spin pattern allows all torques. The light blue arrows represent the rotated spin directions in (b), which are omitted in (c) for clarity.}
    \label{fig:Schm}
\end{figure}

In the collinear magnetic pattern [Fig. \ref{fig:Schm}(a)], the spin-only group is the semidirect product $\mathcal{G}_{\mathrm{SO}}=\mathrm{SO}(2)\rtimes Z_2^{K}$ where $Z_2^{K}=\{E,\bar{C}_2\}$ and $\rm SO(2)$ the arbitrary rotation along the spin axis. Hence, the spin torque is always symmetry-forbidden. This indicates that the NR optically-driven spin torque would not occur in collinear magnets. Actually, one could include relativistic SOC effect to yield finite responses, corresponding to the symmetry reduction from spin group to magnetic group. In this regard, we only focus on noncollinear magnets in the NR regime. 

When the spins locate in the same plane (coplanar) [Fig. \ref{fig:Schm}(b)], the spin-only group becomes $\mathcal{G}_{\mathrm{SO}}=Z_2^{K}$ where the rotation axis is perpendicular to the spin plane (only one direction). Such a group exhibits an effective time-reversal symmetry $\mathcal{T}_{\rm eff}$. We compare the general transformations of OIST responses in conventional ($\mathcal{T}\bm t$ and $\mathcal{PT}$ invariant) antiferromagnets and coplanar NR systems, as tabulated in Table \ref{tab:sym}. One sees that they exhibit different behaviors. While in antiferromagnets the CPL-induced OIST always flip their signs, in $Z_2^{K}$-symmetric coplanar spin configurations it is the light polarization nature that controls the torque direction. In detail, the LPL could induce a torque normal to the spin plane, while the CPL gives a torque lying in the spin plane. Furthermore, if the spin pattern is noncoplanar [Fig. \ref{fig:Schm}(c)], its spin-only group is reduced to identity, $\mathcal{G}_{\mathrm{SO}}=E$, hence no further constraints exist.

\begin{table}[bt]
    \caption{Symmetry constraints of coplanar spin configuration for torque $T$ and OIST \textcolor{black}{responses $\chi$ and $\xi$} in $\mathcal{T}\bm t$ and $\bar{E}\mathcal{T}$ antiferromagnets, and NR $[\bar{C}_{2\perp}||\mathcal{T}]$. Here, the \checkmark~and \xmark~represent symmetry-allowed and forbidden. The $\parallel$ and $\perp$ indicate directions within and normal to the spin plane.}
        \centering
    \begin{tabular*}{\hsize}{@{}@{\extracolsep{\fill}}ccccccc@{}}
    \hline\hline
     Operations & $T_{\perp}$ & $T_{\parallel}$  & $\chi_{\alpha\beta}^{\perp}$ & $\chi_{\alpha\beta}^{\parallel}$ & $\xi_{\alpha\beta}^{\perp}$ & $\xi_{\alpha\beta}^{\parallel}$ \\
    \hline
    $\mathcal{T}\bm t$   &  \checkmark  &  \checkmark  &  \checkmark  &  \checkmark  & \xmark & \xmark \\
    $\bar{E}\mathcal{T}$   & \checkmark  &  \checkmark  &  \checkmark  &  \checkmark  & \xmark & \xmark \\
    $[\bar{C}_{2\perp}||\mathcal{T}]$ & \checkmark  &  \xmark  & \checkmark  &  \xmark  & \xmark & \checkmark \\
    \hline\hline
    \end{tabular*}
    \label{tab:sym}
\end{table}

In detail, the quantum geometric tensors $g^{\alpha\beta}(=g^{\beta\alpha})$ and $\Omega^{\alpha\beta}(=-\Omega^{\beta\alpha})$ follow the lattice transformation operations $R_l$. Without loss of generality, we assume that the light propagates along $z$, and the in-plane quantum metric components $g^{xx}$, $g^{yy}$, and $g^{xy}$ (Berry curvature $\Omega^{xy}$) control the LPL (CPL) effect. Since $g$'s are $\mathcal{T}$ even, they are determined by the \textit{spatial} point group of the system. Of all 32 crystallographic point groups, $g^{xx}$ and $g^{yy}$ are always symmetry-allowed. $g^{xy}$ would flip its sign when the system contains vertical mirror reflection or high order rotations $C_{nz}\,(n\geq 3)$, thus it only survives in $1$, $\bar{1}$, $2$, $m$, and $2/m$ point groups. The Berry curvature $\Omega^{xy}$ is $\mathcal{T}$-odd, which is determined if the system contains effective time-reversal $\mathcal{T}_{\rm eff}$. In general, it is invariant under $C_{nz}$, improper rotation $\bar{C}_{nz}$, $M_{x}\mathcal{T}$ (or other antiunitary vertical mirror reflection such as $M_y\mathcal{T}$, $M_{xy}\mathcal{T}$), $M_z$, and spatial inversion $\bar{E}$, while flips its sign under $C_{nz}\mathcal{T}$, $\bar{C}_{nz}\mathcal{T}$, $M_x$, $M_y$, $M_z\mathcal{T}$, and $\bar{E}\mathcal{T}$.

In most magnetic systems, especially those with $3d$ transition metals, the spin torque can be locally-defined for each magnetic site \textcolor{black}{(note that this is not always the case, especially for itinerant magnets)}. We are interested in the site-dependent NR-OIST ($\chi_{\alpha\beta}^{\gamma,i}$ and $\xi_{\alpha\beta}^{\gamma,i}$ with $i$ the site index) in addition to their entire behavior. For each magnetic site, their site-spin-symmetry operations form a subgroup $\mathcal{G}^i=\{[R_s||R_l^i]\}\leq\mathcal{G}$. Here, $R_l^i:\bm r_i\rightarrow\bm r_i+\bm t$, hence the set $\{R_l^i\}$ forms the spatial site group $\mathcal{H}^i$ (without considering spin configuration). While it is clear that the site-$i$ has a definite spin orientation, $\mathcal{G}^i$ is not simply a spin group of collinear ferromagnets, hence it does not contain $^{m\infty}1$.

The symmetry constraints of \textcolor{black}{$[R_s^i||R_l^i]$} on NR-OIST is determined by if the spin torque and quantum geometric tensors transform in the same manner under $R_s^i$ and $R_l^i$, respectively. The details are given in SM \cite{supp}. Specifically, if the torque $T$ keeps or flips their signs, we have \textcolor{black}{$R_s^{+,i}T_{\gamma}^i=T_{\gamma}^i$ and $R_s^{-,i}T_{\gamma}^i=-T_{\gamma}^i$}. Then, only the $[R_s^{+,i}||R_l^{+,i}]$ and $[R_s^{-,i}||R_l^{-,i}]$ \textcolor{black}{(here, $R_l^{+,i}Q^{\alpha\beta}=Q^{\alpha\beta}$ and $R_l^{-,i}Q^{\alpha\beta}=-Q^{\alpha\beta}$)} could allow the NR-OIST, while it is symmetrically forbidden if \textcolor{black}{$[R_s^{-,i}||R_l^{+,i}]$ or $[R_s^{+,i}||R_l^{-,i}]$} appears in the SPG. \textcolor{black}{Here, $Q^{\alpha\beta}$ is a global quantity, whose local feature is gauge-dependent.} For instance, in the coplanar spin pattern [Fig. \ref{fig:Schm}(b) and Table \ref{tab:sym}], $[\bar{C}_{2z}||\mathcal{T}]$ allows an out-of-plane spin torque under LPL, and an in-plane torque (vertical to each spin) if CPL is applied. For the coplanar spin configurations, we scrutinize all SPGs \cite{Brinkman66PRSA,Liu22PRX} and find there are 90 of them could describe site-SPGs. Furthermore, following this recipe, we list the NR-OIST symmetry constraints for all these site-SPGs (see SM \cite{supp} for the tables and the detailed analysis).

\begin{figure*}[tb]
    \centering
    \includegraphics[width=0.85\textwidth]{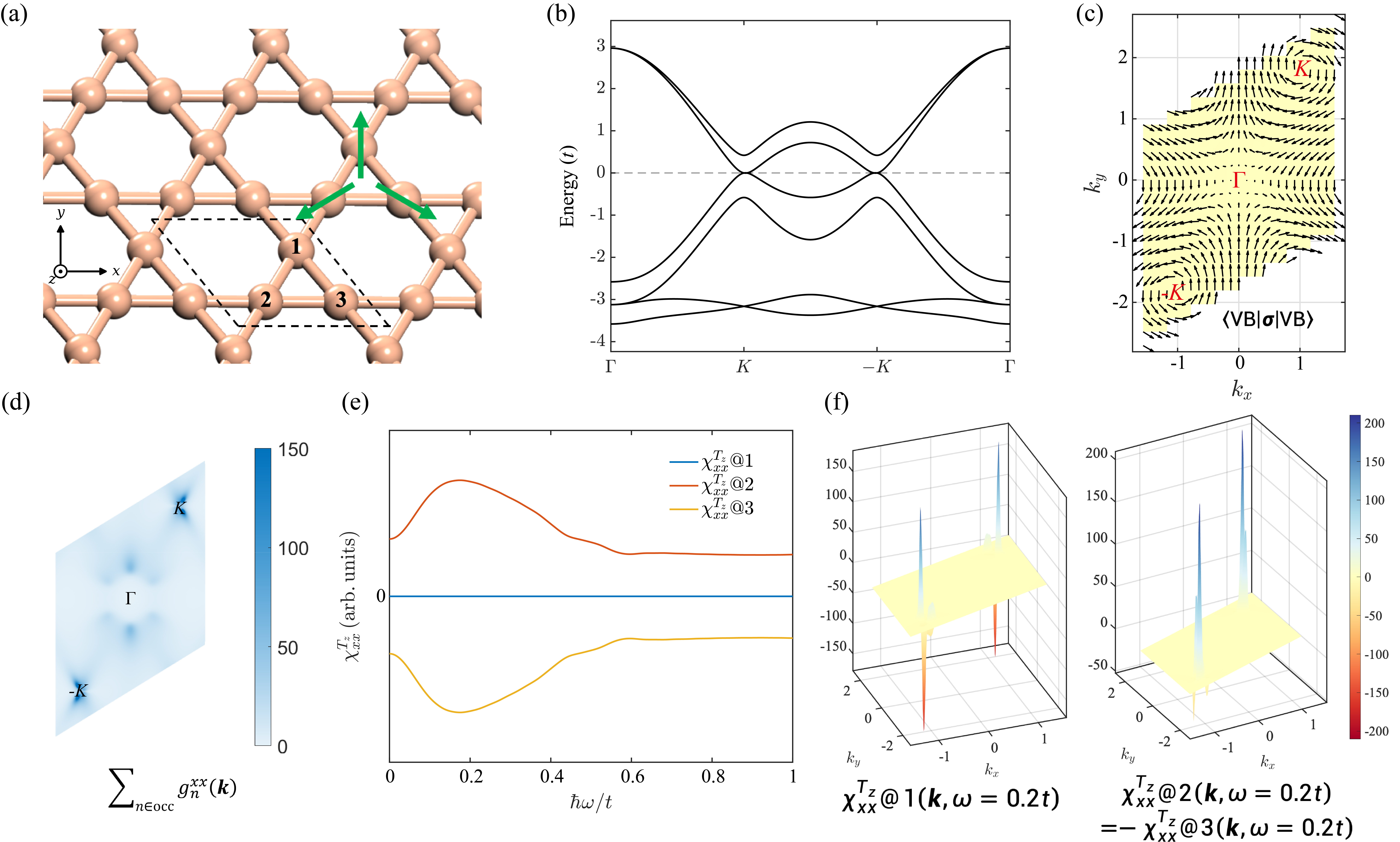}
    \caption{NR-OIST in the Kagom\'e model with $2/3$-filling. (a) Simulation model with $\kappa=1$ spin configuration indicated by the green arrows. (b) Band dispersion along the high-symmetric $k$-path. (c) Spin texture of the top valence band $\langle\mathrm{VB}|\bm\sigma|\mathrm{VB}\rangle$, with the in-plane components indicated by the black arrows, and the out-of-plane component is completely zero (yellow background). (d) Momentum-dependent quantum metric (summed over all valence bands) $\sum_{n\in\mathrm{occ}}g_n^{xx}(\bm k)$. (e) $\chi_{xx}^{T_z}(\omega)$ at the three sites under LPL, and (f) plots their $\bm k$-resolved distribution at a typical incident energy $\hbar\omega=0.2t$.}
    \label{fig:kago}
\end{figure*}

\textit{Tight binding model for Kagom\'e lattice.}---We next illustrate the above analysis using an isotropic Kagom\'e lattice by a simple tight-binding model [Fig. \ref{fig:kago}(a)]. Kagom\'e lattice widely exist in quantum materials, such as $\mathrm{Mn}_3X\,(X=\mathrm{Ge},\mathrm{Sn},\mathrm{Rh},\mathrm{Ir},\mathrm{Pt})$ \cite{Nakatsuji15Nature15723,Nayak16sciadv.1501870,Yang_2017NPJaa5487,Ikhlas17NatPhys4181}, $\mathrm{CsV}_3\mathrm{Sb}_5$ \cite{Wilson24NRMs41578-024-00677-y,Kiesel13PhysRevLett.110.126405}, $\mathrm{Fe}\mathrm{Sn}$ \cite{Kang20NMs41563-019-0531-0} and $\mathrm{Co}\mathrm{Sn}$ \cite{Huang22PhysRevLett.128.096601}, $\mathrm{Co}_3\mathrm{Sn}_2\mathrm{S}_2$ \cite{Liu18NatPhyss41567-018-0234-5}, and $\mathrm{Fe}_3\mathrm{Sn}_2$ \cite{Lin18PhysRevLett.121.096401}. The basic single particle Hamiltonian reads
\begin{equation}\label{eq:Kagomeham}
    H_{\mathrm{K}}=t\sum_{\langle i,j\rangle,\alpha}c_{i,\alpha}^{\dagger}c_{j,\alpha}+J\sum_{i,\alpha\beta}c_{i\alpha}^{\dagger}(\hat{\bm s}_i\cdot\bm\sigma)_{\alpha\beta}c_{i\beta}.
\end{equation}
Here $t$ is the nearest neighbor hopping integral and $J$ measures the exchange energy. $\hat{\bm s}_i=(\cos\varphi_i,\sin\varphi_i,0)$ is the in-plane unit vector along the spin on site-$i$, with $\varphi_i=\varphi_0+i\kappa\frac{2\pi}{3}\,(i=1,2,3)$ is their azimuthal angle. Here $\varphi_0$ denotes a universal rotation angle for all atoms, and then they each differ by $120^{\circ}$. $\kappa=\pm 1$ is the sign of vector spin chirality of Kagom\'e lattice \cite{pradhan2023vector}. When the SOC is absent, the specific value of $\varphi_0$ does not affect the main conclusion. We set $J=0.5t$ in the simulation.

The SPG of this model is $^{3_z}6/^{1}m^{2_y}m^{2_{xy}}m$. With $2/3$-filling, we plot its band dispersion in Fig. \ref{fig:kago}(b), which shows a semimetallic feature at the two valleys $K$ and $-K$. Since the system is centrosymmetric with co-planar spin configuration, the $[\bar{C}_{2z}||\bar{E}\mathcal{T}]$ always present, given the band dispersion $\varepsilon_n(\bm\sigma,\bm k)$ with even-parity in-plane spin texture, under $[\bar{C}_{2z}||\bar{E}\mathcal{T}]\varepsilon_{n}(\sigma_{x,y},\bm k)=\varepsilon_{n}(\sigma_{x,y},\bm k)$ and vanishing out-of-plane spin $[\bar{C}_{2z}||\bar{E}\mathcal{T}]\varepsilon_{n}(\sigma_z,\bm k)=\varepsilon_{n}(-\sigma_z,\bm k)$. This is consistent with our simulation in Fig. \ref{fig:kago}(c). 
For an arbitrary $\varphi_0$, such a persistent spin helix always remains under $[C_{2\hat{\bm s}_1}||M_x]$ and $[C_{2\hat{\bm s}_1}||M_y]$. This is true for both $\kappa=1$ and $\kappa=-1$. One also observes a clockwise and counterclockwise spin texture near the $K$ and $-K$ valleys.

In this particular situation, the Berry curvature is totally forbidden as $[\bar{C}_{2z}||\bar{E}\mathcal{T}]\Omega(\bm k)=-\Omega(\bm k)$. Hence, CPL would not induce any NR-OIST due to the presence of inversion symmetry. We thus focus on the LPL-induced spin torque. As discussed previously, the diagonal quantum metric tensors $g^{xx}$ and $g^{yy}$ are allowed (following $R_l^+$ character). We plot the momentum-resolved $g_n^{xx}(\bm k)$ summed over occupied bands [Fig. \ref{fig:kago}(d)], which exhibit mirror reflection behavior with respect to $k_x=0$ and $k_y=0$, governed by $[C_{2y}||M_y]$ and $[C_{2y}||M_x]$, respectively. However, as for $T_z$, the $[C_{2y}||M_x]$ (or $[C_{2y}||M_y]$) flip its sign (belonging to $R_s^-$), making the total $\chi_{xx}^z(\omega)$ symmetrically forbidden. This is consistent with our simulation results.

\textcolor{black}{For the local site-spin symmetry, all three sites belong to spin group of $^2m^2m^1m$. Their vertical mirror are connected by a $120^\circ$-rotation. Hence, under $x$-LPL, the site-1 is silent (Table S2 in SM \cite{supp}). However, for sites-2 and 3, as their local vertical mirrors are not normal to the global coordinate frame $x$ axis, then one has $\chi^{z,2}_{xx}(\omega)=-\chi^{z,3}_{xx}(\omega)
\neq0$, which are not independent enforced by $[C_{2\hat{s}_1}\|M_x]$.} This can be seen in Fig. \ref{fig:kago}(e). The $\delta$-function in Eq. (\ref{eq:oist}) is represented by the Lorentz-function with a broadening factor of $0.05t$. With this factor, their peaks appear near the $\hbar\omega=0.2t$, which arises from both valleys [Fig. \ref{fig:kago}(f)]. \textcolor{black}{We artificially add a typical Kane-Mele SOC term in the model, which suggests that the exchange controlled NR-OIST is the leading order. It implies that the SOC-related relativistic effects serve as a higher order expansion \cite{supp,Liu25PhysRevX.15.031006}.}

When the light is polarized along the $[110]$ direction, both $[C_{2y}||M_y]$ and $[C_{2y}||M_x]$ now becomes $[R_s^-||R_l^-]$ category for $T_z$. In this regard, one can expect a total NR-OIST over the entire system $\sum_{i=1}^3\chi_{xy}^{z,i}$. This agrees well with our simulation results \textcolor{black}{(see SM \cite{supp})}. \textcolor{black}{It indicates that even though the quantum metric $g^{xy}$ is forbidden (allowed in general $\bm k$ points but canceled in whole BZ by symmetry), one can use this optical approach to indirectly measure its effect. Thus, our NR-OIST provides an additional strategy to detect such a \textit{hidden} quantum geometry nature in noncollinear magnets.}

We briefly evaluate the situation when the spin configuration exhibits a noncoplanar configuration, in which $[\bar{C}_{2z}||\mathcal{T}]$ no longer exists. As shown in SM \cite{supp}, in this case the NR-OIST under CPL irradiation $\xi_{\alpha\beta}^{}$ emerges, consistent with our symmetry arguments.

\begin{figure}[bt]
    \centering
    \includegraphics[width=0.48\textwidth]{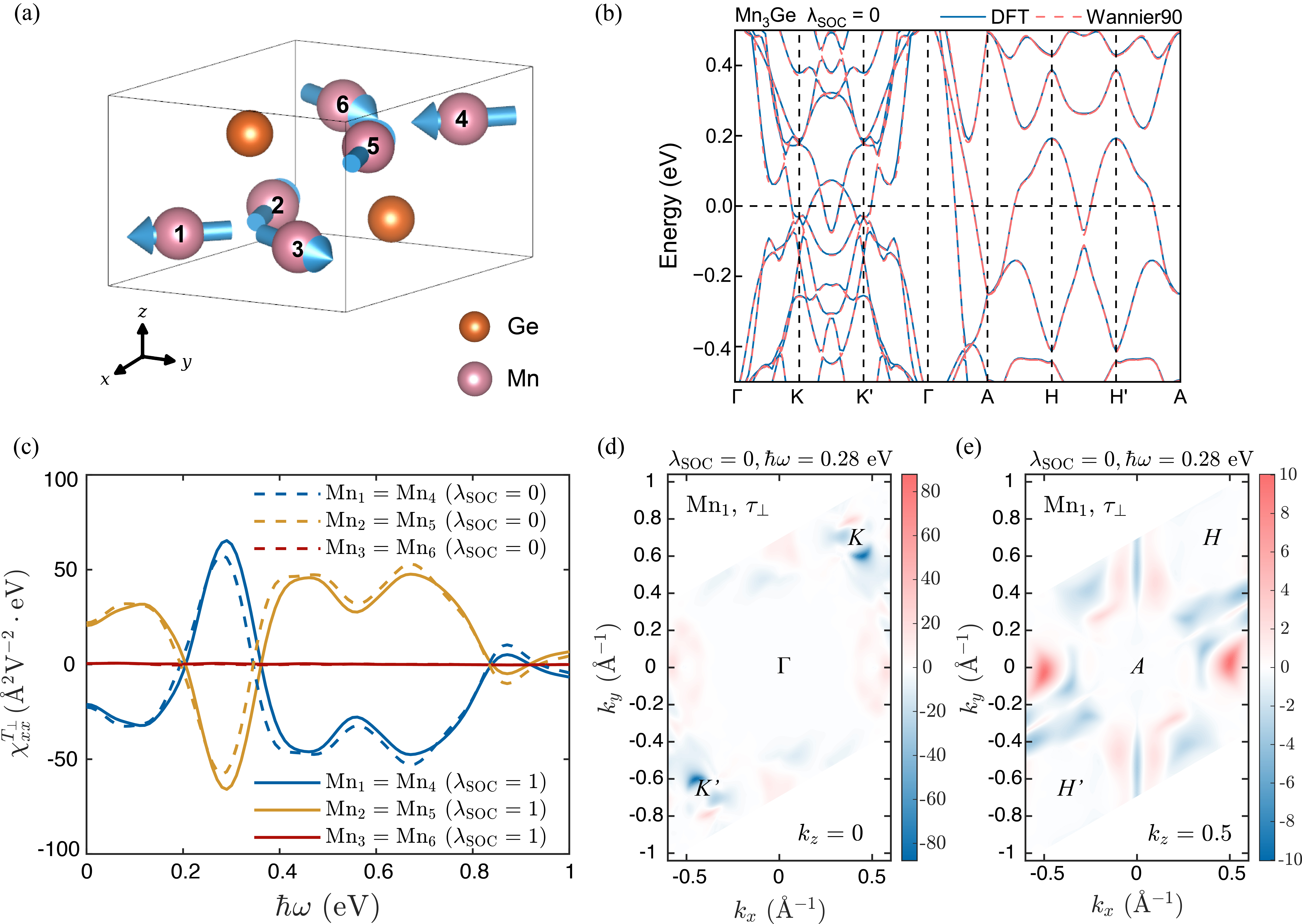}
    \caption{NR-OIST in $\mathrm{Mn}_3\mathrm{Ge}$. (a) Geometric structure in a unit cell with arrows denoting the magnetic pattern. (b) Band dispersion calculated from DFT and fitted by Wannier functions without SOC. (c) $x$-polarized light induced spin torque on the three $\rm Mn$ sites in the NR limit (dashed curves, $\lambda_{\rm so}=0$) and full relativistic (solid curves, $\lambda_{\rm so}=1$) situations. (d) Momentum resolved distribution of NR-OIST at $\rm Mn_1$ on the $k_z=0$ and $k_z=\pi/2c$ slices.}
    \label{fig:Mn3Ge}
\end{figure}

\textit{Realistic material examples.}---We illustrate our theory in the prototypical material $\mathrm{Mn}_3\mathrm{Ge}$ compound [Fig. \ref{fig:Mn3Ge}(a)]. Similar results for $\rm Mn_3Sn$ can be seen in SM \cite{supp}. In both these materials, the $\mathrm{Mn}$ exhibits a Kagom\'e skeleton lattice. From the top view (along $[001]$ direction), they take the all-in-all-out magnetic configuration, which has a right-handed spin chirality with $\kappa=1$. In the NR regime, its nontrivial SPG is $^{3_z}6/^1m^{2_x}m^{2_{xy}}m$.
It contains six $\mathrm{Mn}$ sites in a unit cell. Compared with the Kagom\'e lattice model, the $\mathrm{Mn}$ sites lack of $[E||\bar{E}]$, while the whole system \textcolor{black}{retains} inversion symmetry. We then perform \textit{ab initio} calculations (see computational details in SM \cite{supp}) to compute its NR band dispersion [Fig. \ref{fig:Mn3Ge}(b)]. If full relativistic effect is taken into account \cite{supp}, the degenerate points slightly above the Fermi energy split, while the overall shape of band dispersion does not change too much. This indicates that the relativistic SOC would not greatly affect the NR-OIST responses. \textcolor{black}{This is because the exchange-driven spin splitting ($\sim$0.1--1\,eV) is 1--2 orders of magnitude larger than the SOC-effect, so the quantum geometric tensors and hence the NR-OIST are only perturbatively modified by SOC. We artificially tuned its strength and confirmed this argument \cite{supp}.}

Owing to the combined effective time-reversal symmetry and inversion symmetry, we still focus on the LPL effect. Figure \ref{fig:Mn3Ge}(c) plots our calculated $\chi_{xx}^{{\perp}}(\omega)$ for each $\rm Mn$ sites. One sees that the two $\rm Mn$ planes are equivalent under fractional translation. \textcolor{black}{Again, all six $\rm Mn$ have site-spin group of $^2m{}^2m{}^1m$. The silence of $\rm Mn_3/Mn_6$ arises as their local mirror aligns with the $x$-LPL, not from any higher site symmetry.} We find the dominant peak for other $\rm Mn$ sites reaches $\pm 60\,\mathrm{eV}\cdot\angstrom^2/\mathrm{V}^2$ at an incident energy of $\hbar\omega=0.28\,\mathrm{eV}$. The momentum-resolved contribution at typical slices of $k_z=0$ and $k_z=\pi/2$ are plotted in Fig. \ref{fig:Mn3Ge}(d). One sees that the overall shape follows $\mathcal{T}$-even distribution, consistent with our previous discussions. The dominant contribution arises near the $K$ and $K'$ valleys due to relatively larger joint density of states. The $\chi_{xx}^{\perp}\sim\pm 60\,\mathrm{eV}\cdot\angstrom^2/\mathrm{V}^2$ value indicates that for a moderate light with its electric field strength of $0.1\,\mathrm{V/nm}$ (intensity of $1.3\times 10^{9}\,\rm W/cm^2$), it could exert a torque of $\pm 6\,\rm meV$ on the $\rm Mn$ spin. This torque value can be interpreted as an energy derivative with respect to spin rotation, $T\sim \partial E/\partial\theta$ rather than a conventional mechanical torque. One can estimate the effective magnetic field by $B_{\rm eff}=T/m_{\rm Mn}$, giving a staggered magnetic field of $30\,\rm T$ when $\rm Mn$ magnetic moment taken as $\sim 3\,\mu_B$. This indicates a sizable local perturbation comparable to experimentally accessible magnetic fields. In the energy landscape, one could approximately expand the energy near equilibrium as $E(\theta)=\frac{1}{2}\mathcal{K}\theta^2$, where $\mathcal{K}$ is an effective magnetic stiffness primarily set by exchange interactions. Notably, the induced variation $\delta m$ is transverse to the equilibrium magnetization direction, so the torque drives a rotation of these moments without changing its magnitude. According to \textcolor{black}{our estimates, the dominant exchange scale for $\rm Mn$ lies on the order of $100\,\rm meV$ \cite{supp}, while the anisotropic energy is order of magnitude smaller}. This is also consistent with previous works \cite{Chen14PhysRevLett.112.017205,Kubler2014EPL}. This implies that the rotation of magnetic moment will result in a canting angle of $\pm 0.1\,\rm rad$. As a second order nonlinear optical effect, such rotation is proportional to electric field magnitude. The opposite values for $\rm Mn_1$ ($\rm Mn_4$) and $\rm Mn_2$ ($\rm Mn_5$) suggest that they tend to align in the parallel manner with $\rm Mn_3$ ($\rm Mn_6$). \textcolor{black}{Thus, the NR-OIST would slightly cant the $120^{\circ}$ spin configuration toward collinear pattern, leading to nonzero total spin moments that can be detected in experiments. After light is turned off, it would gradually restore the equilibrium state over a few picoseconds (see discussions and estimate in SM \cite{supp}).} Overall, this magnitude signifies a moderate but non-negligible spin reorientation.

The CPL effect, as discussed previously, would be silent under NR-limit. However, we find that it arises under finite SOC strength, implying a relativistic nature of CPL-induced spin torque, see SM \cite{supp}. For instance, under a left-handed light, $\xi_{\sigma^+}^{{\parallel}}$ reaches a negative peak of $-4\,\mathrm{eV}\cdot\angstrom^2/\mathrm{V}^2$ at an incident energy of $\hbar\omega=0.49\,\mathrm{eV}$, being much smaller than NR-OIST under LPL.

\textit{Conclusion.}---In summary, we perform a group-theoretical and numerical simulation combined study to show how optical field exerts spin torques in noncollinear magnets in the NR-limit. Such a nonlinear optical process could help uncover the quantum geometric nature of Bloch wavefunctions. Furthermore, we illustrate our theory using a Kagom\'e lattice model and realistic $\rm Mn_3Ge/Mn_3Sn$ systems, unraveling observable OIST effect that \textcolor{black}{mimics} a giant effective magnetic field. \textcolor{black}{Our predictions can be applied in a broad family of noncollinear magnets, such as $\mathrm{Mn}_3X$ ($X = \rm Ir, Pt, Rh$) \cite{Nakatsuji15Nature15723,Nayak16sciadv.1501870,Yang_2017NPJaa5487,Ikhlas17NatPhys4181}, antiperovskite $\mathrm{Mn}_3A\mathrm{N}$ ($A = \rm Ni, Ga, Zn$) \cite{PhysRevMaterials.3.044409,PhysRevMaterials.3.094409}, and $\rm Fe_3Sn_2$ \cite{Lin18PhysRevLett.121.096401}, etc.} Our work opens the route to noncontacting optical control of magnetic configurations in the NR limit.

\begin{acknowledgments}
\textcolor{black}{\textit{Acknowledgment.}} This work is supported by the National Natural Science Foundation of China (NSFC) under Grant No. 12374065. The authors thank the support of the 111 Project (B25007). The HPC platform of Xi'an Jiaotong University is also acknowledged where the computations are performed.
\end{acknowledgments}

\textit{Data availability.} The data that support the findings of this article are not publicly available yet. The data are available from the authors upon reasonable request.

\bibliography{Mainbib}

\onecolumngrid
\vspace{1em}
\begin{center}
	\textbf{\large \textcolor{black}{End Matter}}
\end{center}
\vspace{1em}
\twocolumngrid

\textcolor{black}{
We briefly provide the density matrix approach for the OIST responses, in the length gauge framework. The detail algebraic derivations can be seen in SM \cite{supp}. The expectation of value of spin torque along $a$-direction is (we omit the site index $i$ for simplicity, and set $e=\hbar=1$)
\begin{equation}
    T^{a}=\mathrm{Tr}(\hat{\rho}{T}^{a})=\sum_{n,m}\int[d\bm k]T_{nm}^{a}\hat{\rho}_{mn}.
\end{equation}
The density matrix under alternating electric fields can be expanded according to quantum Liouville equation \cite{Aversa95PhysRevB.52.14636,Sipe20PRB,Liu25PRLnsky-zmzp}}
\textcolor{black}{
\begin{equation}
    \hat{\rho}_{mn}=\sum_{l=0}^{\infty}\hat{\rho}_{mn}^{(l)}=\hat{\rho}_{mn}^{(0)}+\hat{\rho}_{mn}^{(1)}+\hat{\rho}_{mn}^{(2)}+\cdots,
\end{equation}
where $\hat{\rho}_{mn}^{(0)}=\delta_{mn}f_n$ with $f_n$ the equilibrium Fermi-Dirac distribution function. The second order modification of density matrix is}
\textcolor{black}{
\begin{equation}
    \begin{split}
        \hat{\rho}_{mn}^{(2)}=&\frac{i\hat{\rho}_{mn;c}^{b,(1)}+\sum_l(r_{ml}^c\hat{\rho}_{ln}^{b,(1)}-\hat{\rho}_{ml}^{b,(1)}r_{ln}^c)}{\tilde{\omega}_{\Sigma}-\omega_{mn}}E_{\omega_{\beta}}^bE_{\omega_{\gamma}}^ce^{-i\tilde{\omega}_{\Sigma}t} \\
        =&\frac{1}{\tilde{\omega}_{\Sigma}-\omega_{mn}}\left\{-\left(\frac{\hat{\rho}_{mn;b}^{(0)}}{\tilde{\omega}_{\beta}-\omega_{mn}}\right)_{;c}+i\left(\frac{\hat{q}_{mn}^{b,(0)}}{\tilde{\omega}_{\beta}-\omega_{mn}}\right)_{;c}\right. \\
        +&\left. \sum_l\left[r_{ml}^c\left(\frac{i\hat{\rho}_{ln;b}^{(0)}+\hat{q}_{ln}^{b,(0)}}{\tilde{\omega}_{\beta}-\omega_{ln}}\right)-\left(\frac{i\hat{\rho}_{ml;b}^{(0)}+\hat{q}_{ml}^{b,(0)}}{\tilde{\omega}_{\beta}-\omega_{ml}}\right)r_{ln}^b\right]\right\} \\
        \times & E_{\omega_{\beta}}^bE_{\omega_{\gamma}}^ce^{-i\tilde{\omega}_{\Sigma}t}.
    \end{split}
\end{equation}
}

\textcolor{black}{
In these equations, we adopt following conventional notations, $\tilde{\omega}_{\beta}=\omega_{\beta}+i\eta$, with $\eta$ being a small real number to adiabatically switching on the the field which numerically incorporate the broadening factor. $\omega_{mn}=\omega_m-\omega_n$ is the eigenfrequency difference between Bloch state $m$ and $n$. The covariant derivative $\mathcal{D}^b\hat{\mathcal{O}}_{mn}=\partial_{k_b}\hat{\mathcal{O}}_{mn}-i(\xi_{mm}^b-\xi_{nn}^b)\hat{\mathcal{O}}_{mn}$ is denoted as $\hat{\mathcal{O}}_{;b}$ with $\xi_{nn}^b=i\langle n|\partial_{k_b}|n\rangle$ the intraband Berry connection. The second order frequency arises from $\omega_{\Sigma}=\omega_{\beta}+\omega_{\gamma}$, and we use a phenomenological universal carrier lifetime $\tau=1/\eta$ to represent the disorder, scattering, electron-electron interaction, impurity effects, etc.}

\textcolor{black}{
Usually we partition the density matrix into two parts, namely, the interband off-diagonal $m\neq n$ (OD) and intraband diagonal $m=n$ (D) contributions. They are
\begin{widetext}
\begin{equation}\label{eq:rho2}
    \begin{split}
        \hat{\rho}_{mn}^{(2),\mathrm{OD}}=&\frac{1}{\tilde{\omega}_{\Sigma}-\omega_{mn}}\left[-\left(\frac{\hat{\rho}_{mn;b}^{(0)}}{\tilde{\omega}_{\beta}-\omega_{mn}}\right)_{;c}+i\left(\frac{\hat{q}_{mn}^{b,(0)}}{\tilde{\omega}_{\beta}-\omega_{mn}}\right)_{;c}\right]E_{\omega_{\beta}}^bE_{\omega_{\gamma}}^ce^{-i\tilde{\omega}_{\Sigma}t} \\
        +&\frac{1}{\tilde{\omega}_{\Sigma}-\omega_{mn}}\sum_l\left[r_{ml}^c\left(\frac{i\hat{\rho}_{ln;b}^{(0)}+\hat{q}_{ln}^{b,(0)}}{\tilde{\omega}_{\beta}-\omega_{ln}}\right)-\left(\frac{i\hat{\rho}_{ml;b}^{(0)}+\hat{q}_{ml}^{b,(0)}}{\tilde{\omega}_{\beta}-\omega_{ml}}\right)r_{ln}^c\right]E_{\omega_{\beta}}^bE_{\omega_{\gamma}}^ce^{-i\tilde{\omega}_{\Sigma}t},\\
        \hat{\rho}_{nn}^{(2),\mathrm{D}}=&\frac{1}{\tilde{\omega}_{\beta}\tilde{\omega}_{\Sigma}}\left(-\partial_{k_b}\partial_{k_c}\hat{\rho}_{nn}^{(0)}+i\partial_{k_c}\hat{q}_{nn}^{b,(0)}\right)E_{\omega_{\beta}}^bE_{\omega_{\gamma}}^ce^{-i\tilde{\omega}_{\Sigma}t} \\
        +&\frac{1}{\tilde{\omega}_{\Sigma}}\sum_l\left[r_{nl}^c\left(\frac{i\hat{\rho}_{ln;b}^{(0)}+\hat{q}_{ln}^{b,(0)}}{\tilde{\omega}_{\beta}-\omega_{ln}}\right)-\left(\frac{i\hat{\rho}_{nl;b}^{(0)}+\hat{q}_{nl}^{b,(0)}}{\tilde{\omega}_{\beta}-\omega_{nl}}\right)r_{ln}^c\right]E_{\omega_{\beta}}^bE_{\omega_{\gamma}}^ce^{-i\tilde{\omega}_{\Sigma}t}
    \end{split}
\end{equation}
\end{widetext}
}

\textcolor{black}{
In the nonrelativistic regime with spin-orbit coupling $H_{\rm so}$ omitted, the nondiagonal covariant derivative is $T_{nm;a}^b=(i\omega_{nm}\sigma_{nm}^b)_{;a}=i\Delta_{nm}^a\sigma_{nm}^b+i\omega_{nm}\sigma_{nm;a}^b=i\Delta_{nm}^a\sigma_{nm}^b-\omega_{nm}[\sum_l(r_{ml}^a\sigma_{ln}^b-\sigma_{ml}^br_{nl}^a)],\,(n\neq m)$. This is because the total Hamiltonian is $H=H_0+H_{\rm ex}$ with $H_0$ independent on spin operators. We then multiply the torque matrix element with the off-diagonal and diagonal density matrix, with $\omega_{\beta}=-\omega_{\gamma}$ and $\omega_{\Sigma}=0$ for second order rectification process. After some straightforward algebraic derivation \cite{supp}, we have}
\textcolor{black}{
\begin{equation}
    \begin{split}
        T_{nm}^{a} \hat{\rho}_{mn}^{(2),\text{OD}} =\left[\frac{r_{mn}^{b} f_{nm} (\sigma_{nm;c}^{a}-\sigma_{mn;c}^{a})}{\omega_{\beta} - \omega_{mn} + i\eta}\right]E_{\omega_{\beta}}^bE_{\omega_{\gamma}}^ce^{-i\tilde{\omega}_{\Sigma}t}
    \end{split}
\end{equation}
}
\textcolor{black}{
The sum rule of covariant derivative for Pauli matrix is 
\begin{equation}
    \sigma_{nm;a}^b=i\sum_l(r_{nl}^a\sigma_{lm}^b-\sigma_{nl}^br_{lm}^a).
\end{equation}
In order to obtain the final expression, we symmetrize the $b\beta$ and $c\gamma$. Then, the OIST response functions are
\begin{equation}
    \begin{split}
        \chi_{bc}^{T_a,\mathrm{OD}}(\omega)=&i\pi\int[d\bm k]f_{nm}(r_{mn}^b\sigma_{nm;c}^a-r_{nm}^c\sigma_{mn;b}^a)\\
    &\times\delta(\omega-\omega_{mn}),
    \end{split}
\end{equation}
and
\begin{equation}
    \begin{split}
        \xi_{bc}^{T_a,\mathrm{OD}}(\omega)=&\pi\int[d\bm k]f_{nm}(r_{mn}^b\sigma_{nm;c}^a+r_{nm}^c\sigma_{mn;b}^a)\\
    &\times\delta(\omega-\omega_{mn}).
    \end{split}
\end{equation}
They correspond to the linearly and circularly polarized light effects, respectively.}

\textcolor{black}{
The diagonal term of the density matrix contributions are
\begin{widetext}
\begin{equation}
    \begin{split}
        T_{nn}^a\hat{\rho}_{nn}^{(2),\rm D}=
        &\left[\frac{1}{\tilde{\omega}_{\Sigma}} f_{nm} \left( \frac{r_{mn}^{c} \Delta T_{mn}^{a} r_{nm}^{b}}{\omega_{\beta} - \omega_{nm} + i\eta} - \frac{r_{nm}^{c} \Delta T_{nm}^{a} r_{mn}^{b}}{\omega_{\beta} - \omega_{mn} + i\eta} \right) \right]\times E_{\omega_{\beta}}^bE_{\omega_{\gamma}}^ce^{-i\tilde{\omega}_{\Sigma}t}.     
    \end{split}
\end{equation}
\end{widetext}
Similarly, we symmetrize the $b\beta$ and $c\gamma$ to obtain the final expression. The intraband diagonal susceptibilities are
\begin{equation}
    \begin{split}
        \chi_{bc}^{T_a,\mathrm{D}}(\omega)=&\tau\pi\int[d\bm k]f_{nm}\Delta T_{mn}^a(r_{nm}^br_{mn}^c+r_{nm}^cr_{mn}^b)\\
        &\times\delta(\omega-\omega_{mn}),
    \end{split}
\end{equation}
and
\begin{equation}
    \begin{split}
        \xi_{bc}^{T_a,\mathrm{D}}(\omega)=&-i\tau\pi\int[d\bm k]f_{nm}\Delta T_{mn}^a(r_{nm}^br_{mn}^c-r_{nm}^cr_{mn}^b)\\
        &\times\delta(\omega-\omega_{mn}).
    \end{split}
\end{equation}
}
\textcolor{black}{Here, we denote $\Delta T_{nm}^a=T_{nn}^a-T_{mm}^a$ referring to the band difference of the spin torque between $n$ and $m$. These are the main equations used in the main text, reflecting their quantum geometry $Q=g+\frac{i}{2}\Omega$ characters. Note that actually the diagonal (intraband) terms grows linearly with time and saturates at effective lifetime $\tau$. In order to compare them with their off-diagonal interband terms, we take a typical and conservative value here.}

\end{document}